\documentclass[twocolumn,prb,showpacs,floatfix]{revtex4}
\usepackage{graphicx}
\usepackage{bm}

\begin{document}

\title{\textbf{Charge disproportionation and collinear magnetic order \\
in the frustrated triangular antiferromagnet AgNiO$_2$}}
\author{E. Wawrzy\'{n}ska$^{1}$, R. Coldea$^{1}$,
E.M. Wheeler$^{2,3}$, T. S\"{o}rgel$^{4}$, M. Jansen$^{4}$, R.M.
Ibberson$^{5}$, P.G. Radaelli$^{5,6}$, M.M. Koza$^{3}$}
\affiliation{$^1$H.H. Wills Physics Laboratory, University of
Bristol, Tyndall Avenue, Bristol, BS8 1TL, United Kingdom \\
$^2$Clarendon Laboratory, University of Oxford, Parks Road, Oxford
OX1 3PU, United Kingdom\\
 $^3$Institut Laue-Langevin, BP 156, 38042 Grenoble Cedex 9, France\\
 $^4$Max-Planck Institut f\"{u}r Festk\"{o}rperforschung,
 Heisenbergstrasse 1, D-70569 Stuttgart, Germany \\
 $^5$ISIS Facility, Rutherford Appleton Laboratory, Chilton, Didcot
OX11 0QX, United Kingdom\\
 $^6$Dept. of Physics and Astronomy, University College London,
Gower Street, London WC1E 6BT, United Kingdom}
\date{\today}

\begin{abstract}
We report a high-resolution neutron diffraction study of the
crystal and magnetic structure of the orbitally-degenerate
frustrated metallic magnet AgNiO$_2$. At high temperatures the
structure is hexagonal with a single crystallographic Ni site,
low-spin Ni$^{3+}$ with spin-1/2 and two-fold orbital degeneracy,
arranged in an antiferromagnetic triangular lattice with
frustrated spin and orbital order. A structural transition occurs
upon cooling below 365 K to a tripled hexagonal unit cell
containing three crystallographically-distinct Ni sites with
expanded and contracted NiO$_6$ octahedra, naturally explained by
spontaneous charge order on the Ni triangular layers. No
Jahn-Teller distortions occur, suggesting that charge order occurs
in order to lift the orbital degeneracy. Symmetry analysis of the
inferred Ni charge order pattern and the observed oxygen
displacement pattern suggests that the transition could be
mediated by charge fluctuations at the Ni sites coupled to a soft
oxygen optical phonon breathing mode. At low temperatures the
electron-rich Ni sublattice (assigned to a valence close to
Ni$^{2+}$ with $S=1$) orders magnetically into a collinear stripe
structure of ferromagnetic rows ordered antiferromagnetically in
the triangular planes. We discuss the stability of this uncommon
spin order pattern in the context of an easy-axis triangular
antiferromagnet with additional weak second neighbor interactions
and interlayer couplings.
\end{abstract}

\maketitle

\section{Introduction}
\label{sec_introduction}
Two-dimensional frustrated quantum magnets have attracted a wide
interest theoretically\cite{review:theory} and
experimentally\cite{review:exp} as possible candidates to display
strong fluctuations that could potentially stabilize
unconventional ordered phases,\cite{khomskii}
spin-liquid\cite{nature:sl}, or orbital-liquid\cite{vernay}
states. Low-spin triangular-lattice antiferromagnets are canonical
frustrated models and the delafossite AgNiO$_2$ with Ni ions
arranged in well-separated triangular lattices has been proposed
to show frustration effects both in the magnetic as well as in the
orbital sector. Based on magnetic susceptibility measurements Shin
et al.\cite{shin} proposed that Ni ions are in the low-spin
configuration Ni$^{3+}$ ($t_{2g}^6e_g^1$) with $S$=1/2, coupled by
dominant in-plane frustrated antiferromagnetic interactions, and
from x-ray measurements they proposed a high-symmetry crystal
structure where each Ni ion has an unpaired electron in a
doubly-degenerate $e_g$ orbital. The cooperative orbital order is
strongly frustrated because the orbital interactions have a strong
bond-directional dependence\cite{orbital:exchange} favoring
different orbitals for pairs of Ni ions along the three different
directions in the triangular lattice and this leads to a large
manifold of degenerate mean-field states. The ground state in this
spin-orbital problem is still highly debated theoretically with
proposals ranging from orbital liquids to non-trivial forms of
orbital order depending on fine details of the interactions.
\cite{khomskii,vernay,linio2} Among the experimentally explored
candidate spin-1/2 materials to display this physics are NaTiO$_2$
(non-magnetic after structural transition at low
temperatures),\cite{natio2} NaNiO$_2$ (quasi-2D spin-1/2
ferromagnet with ferro-distortive orbital order)\cite{nanio2}, and
LiNiO$_2$ (no long-range magnetic or orbital order, but evidence
for local Jahn-Teller distortions, difficulty in preparing pure
stoichiometric samples).\cite{linio2} The triangular magnet
AgNiO$_2$ and the two-silver-layer version Ag$_2$NiO$_2$ (Refs.
\onlinecite{ag2nio2, soergel07}) are relatively unexplored
experimentally and promise to be rather different from the
above-mentioned systems as they both show dominant {\em
antiferromagnetic} interactions,\cite{ag2nio2_khomskii} thus
potentially displaying frustrated magnetism.

\begin{figure}[tb]
\begin{center}
  \includegraphics[width=8cm,bbllx=55,bblly=227,bburx=539,
  bbury=568,angle=0,clip=] {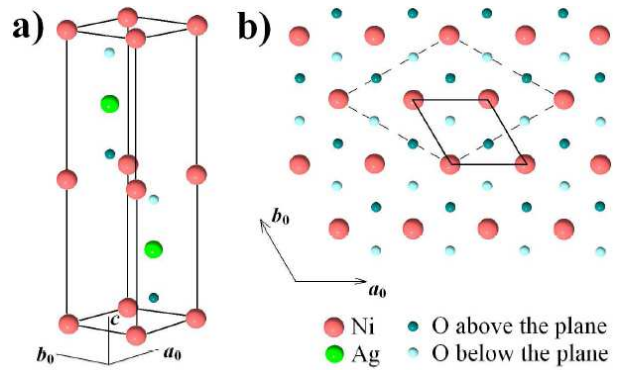}
  \caption{\label{fig_struct_ideal} (color online)
   Nominal crystal structure of 2H-AgNiO$_2$
   deduced from X-ray measurements in Ref.\ \onlinecite{soergel}
   (space group P$6_3/mmc$-D$_{6\rm{h}}^4$). (a) There are two NiO$_2$ layers
   per unit cell related by a mirror plane reflection through
   the Ag$^+$ layer at $z=1/4$. (b) Basal plane showing the triangular
   network of Ni ions (large red balls) coordinated by oxygens (small blue balls).
   Thick solid line contour shows the unit cell and dashed line shows the
   unit cell tripling in the distorted structure.}
\end{center}
\end{figure}

AgNiO$_2$ is part of the large family of delafossite
$A^{+}M^{3+}$O$_2$ materials often studied as candidate
two-dimensional frustrated magnets because the transition metal ion
($M$) sits at the vertices of a triangular lattice in the basal
plane, made up of a network of edge-sharing $M$O$_6$ octahedra. Like
most delafossites, it occurs in two structural polytypes which
differ in the way the NiO$_2$ layers are stacked along the $c$-axis:
a 3-stage structure, where successive layers are in the same
orientation but have an in-plane offset with Ni ions forming a
3-stage staircase along $c$-axis (3R polytype, rhombohedral space
group R$\overline{3}m$, Refs. \onlinecite{agnio2early,shin}), or a
2-stage structure, where successive layers are stacked right on top
of each other but are rotated by 180$^{\circ}$ (2H polytype,
hexagonal space group P$6_3/mmc$ shown in Fig.\
\ref{fig_struct_ideal}(a), Ref. \onlinecite{soergel}). All
measurements reported here have been made on the less-studied
2-stage polytype, the so called 2H-AgNiO$_2$, recently synthesized
using high-oxygen pressure techniques.\cite{soergel} This polytype
has metallic-like conductivity from 300~K down to low temperatures
and the susceptibility indicates antiferromagnetic order near 20
K,\cite{soergel} but the magnetic structure has not been determined
up to now.

In the ideal crystal structure of both 3R and 2H polytypes of
AgNiO$_2$ Ni ions have both spin and orbital degrees of freedom.
The local crystal field is octahedral near-cubic and in the case
of strong crystal field proposed herein\cite{shin,soergel} the
electronic state of Ni$^{3+}$ (3$d^7$) is the low-spin state
$t_{2g}^6e_g^1$ with one unpaired electron (spin-1/2) in the upper
$e_g$ level. A small trigonal distortion present in the crystal
structure due to squashing of the NiO$_6$ octahedra along the
$c$-axis changes the detailed wavefunctions of the orbital states
but does not lift the two-fold degeneracy of the upper $e_g$ level
because it preserves a local three-fold symmetry rotation axis
along $c$. Each Ni ion has a tendency to locally distort the
environment to lower its orbital energy due to the Jahn-Teller
effect, however the cooperative orbital order on the triangular
lattice is frustrated as the orbital exchange favors occupation of
different orbitals for pairs of Ni ions along the three different
in-plane directions. Such systems are susceptible to form an
orbital liquid state at low temperatures or to have the orbital
degeneracy lifted by structural distortions.

In measurements reported here we find evidence for a weak
structural modulation in 2H-AgNiO$_2$ leading to a tripling of the
unit cell in the hexagonal basal plane. This can be naturally
explained by charge disproportionation on the Ni sites into three
sublattices, which we propose occurs in order to lift the orbital
degeneracy of the Ni$^{3+}$ ions. This physics is in sharp
contrast to the insulator NaNiO$_2$ where the orbital degeneracy
is lifted by Jahn-Teller orbital order leading to a monoclinic
crystal structure.\cite{nanio2} We attribute this difference to
the fact that 2H-AgNiO$_2$ being metallic charge transfer can be
an energetically more favourable mechanism to lift the orbital
degeneracy compared to local Jahn-Teller distortions found in more
localized systems.

At low temperatures the electron-rich Ni sublattice (attributed to
a valence close to Ni$^{2+}$ with spin $S=1$) orders magnetically
in a collinear stripe structure with spins pointing along the
$c$-axis and arranged in alternating ferromagnetic rows in the
triangular plane. This magnetic structure cannot be explained at
the mean-field level by a minimal spin model on a triangular
lattice containing only nearest-neighbor antiferromagnetic
exchange and easy-axis anisotropy, and we propose that it is
stabilized by additional weak second neighbor antiferromagnetic
in-plane interactions and/or weak ferromagnetic interlayer
couplings.

The rest of the paper is organized as follows. Next Section
(\ref{sec_exp}) presents the experimental details of the neutron,
X-ray, susceptibility and specific heat measurements. Diffraction
measurements of the room-temperature crystal structure are
presented and analyzed in Sec.\ \ref{sec_crystallography} where a
lower-symmetry space group is proposed to accommodate the observed
structural modulation. A transition to the high-symmetry,
undistorted crystal structure is observed upon heating to high
temperatures and this is discussed in Sec.\ \ref{sec_hight}
followed by a symmetry analysis of the structural distortion in
terms of symmetry-allowed basis vectors in Sec.\ \ref{sec_sym}.
This is used to propose that the mechanism for the structural
distortion is charge fluctuations at the Ni site coupled with a
soft zone-boundary optical phonon involving oxygen breathing modes
(further calculations using co-representation symmetry analysis to
uniquely determine the distorted space group are presented in
Appendix B). Susceptibility and specific heat measurements are
shown in Sec.\ \ref{sec_sus}, followed by measurements of the
magnetic structure analyzed in terms of symmetry-allowed basis
vectors and discussed in terms of a minimal Hamiltonian containing
exchange and easy-axis anisotropy (Sec.\ \ref{sec_mag_order}).
Finally, the results are summarized and discussed in Sec.\
\ref{sec_conclusions}. For completeness we include in Appendix
\ref{app_A} a list of the measured supercell and magnetic
structure factors. A partial account of the results describing the
room-temperature crystal structure and low-temperature magnetic
order has been reported in ref.\ \onlinecite{short_paper}.

\section{Experimental details}
\label{sec_exp}
Powder samples of the 2H-AgNiO$_2$ ($<1\%$ admixture of the 3R
polytype) were prepared from Ag$_2$O and Ni(OH)$_2$ using high
oxygen pressures (130 MPa) as described in Ref.
\onlinecite{soergel}. Neutron diffraction patterns to probe the
crystal and magnetic structure were collected using the
high-resolution back-scattering time-of-flight diffractometers
OSIRIS (0.65$<Q<$6\ \AA$^{-1}$) and HRPD (2$<Q<$9\ \AA$^{-1}$) at
the ISIS Facility of the Rutherford Appleton Laboratory in the UK.
Preliminary measurements were also performed using GEM at ISIS and
the monochromatic neutron diffractometer D1B at the Institute
Laue-Langevin in France. The magnetic order parameter was obtained
from elastic neutron scattering measurements using the
direct-geometry time-of-flight spectrometer IN6 and the
temperature-dependence of the lattice constants was also measured
on D2B, both at the ILL. X-ray powder diffraction measurements to
help solve the crystal structure at room temperature were made
using a Philips X'pert diffractometer ($\lambda_{\rm Cu
K_{\alpha}}=1.54$ \AA). Structural and magnetic refinement was
made using the FullProf programme.\cite{fullprof} Susceptibility
measurements were made using a SQUID magnetometer (Quantum Design
MPMS) and specific heat data was collected on a pressed powder
pellet using a Quantum Design PPMS system.

\begin{figure*}[tb]
\begin{center}
  \includegraphics[width=16cm,bbllx=54,bblly=308,bburx=540,
  bbury=536,angle=0,clip=] {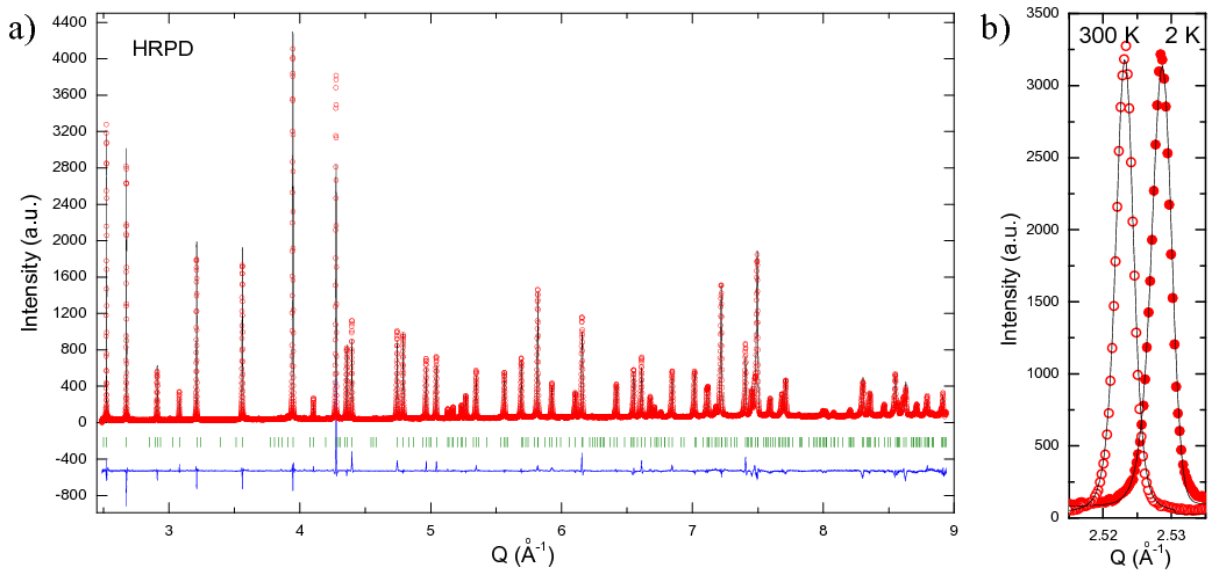}
  \caption{\label{fig_hrpd} (color online)
  (a) Room temperature (300~K) neutron diffraction pattern obtained using
  HRPD (resolution $\Delta Q/Q\sim10^{-3}$, 15 hours counting
  on an 11 g powder sample). The solid curve through data
  points is a fit to the distorted P6$_3 2 2$ space group.
  Vertical bars indicate Bragg peak positions and
  the bottom curve shows the difference between the fit and
  the data.
  Zoomed-in regions showing weak
  supercell peaks are plotted in Fig.\ \ref{fig_hrpd_zoom}.
  (b) Detail of the (111) peak lineshape expected to
  split in the case of a structural distortion
  to an orthorhombic or monoclinic structure. No splitting could
  be detected, and the lineshape at both 300~K (open symbols)
  and 2~K (filled circles) could be well described by a resolution-convolved
  profile for the hexagonal P$6_3 2 2$ space group (solid lines) with
  lattice parameters adjusted for thermal contraction upon cooling.}
\end{center}
\end{figure*}

\section{Crystal structure}
\label{sec_crystallography}

\begin{table}[tb]
\begin{center}
\caption{\label{tab_unitcell} Lattice parameters and atomic
positions in the unit cell in the ideal (P$6_3/mmc$) and the
distorted (P$6_3 2 2$) structural space groups at 300 K. Oxygen
positions are parameterized by an out-of-plane height $z_{\rm
O}=0.08050(5)$ and an in-plane displacement $\epsilon=0.0133(2)$.
The distorted unit cell is tripled in the hexagonal plane with an
unchanged extent along the $c$-axis, but the origin is shifted by
${\bm c}/4$ such that the two NiO$_2$ layers appear now at $z=1/4$
and $3/4$. Throughout this paper we use different symbols, $a_0$
and $a=\sqrt{3}a_0$, to denote the hexagonal lattice parameter of
the ideal and distorted structures, respectively.}
\begin{tabular}{l|l} \hline
 P$6_3/mmc$ (no. 194) & P$6_3 2 2$ (no. 182) \\
\hline
$a_0 = 2.93919(5)$ \AA & $a=5.0908(1)$ \AA \\
$c= 12.2498(1)$ \AA & $c=12.2498(1)$ \AA\\
\hline
\begin{tabular}{lll}
Atom & Site & $(x,y,z)$\\
\hline
   ~~&        ~~&              \\
Ni ~~&   2$a$ ~~& $(0,0,0)$  \\
   ~~&        ~~&              \\
Ag ~~&   2$c$ ~~& $(\frac{2}{3},\frac{1}{3},\frac{1}{4})$\\
O ~~&   4$f$ ~~& $(\frac{2}{3},\frac{1}{3},z_{\rm O})$\\
\end{tabular}
&
\begin{tabular}{lll}
Atom & Site & $(x,y,z)$\\
\hline
Ni1 ~~&   2$c$ ~~& $(\frac{1}{3},\frac{2}{3},\frac{1}{4})$ \\
Ni2 ~~&   2$b$   ~~&  $(0,0,\frac{1}{4})$   \\
Ni3 ~~&   2$d$ ~~& $(\frac{1}{3},\frac{2}{3},\frac{3}{4})$  \\
Ag ~~&   6$g$   ~~& $(\frac{2}{3},0,0)$\\
O  ~~&   12$i$ ~~& $(\frac{1}{3},\epsilon,\frac{1}{4}+z_{\rm O})$
\\[2pt]
\end{tabular}
\\
\hline
\end{tabular}
\end{center}
\end{table}

\begin{figure*}[tb]
\begin{center}
  \includegraphics[width=16cm,bbllx=55,bblly=280,bburx=539,
  bbury=562,angle=0,clip=] {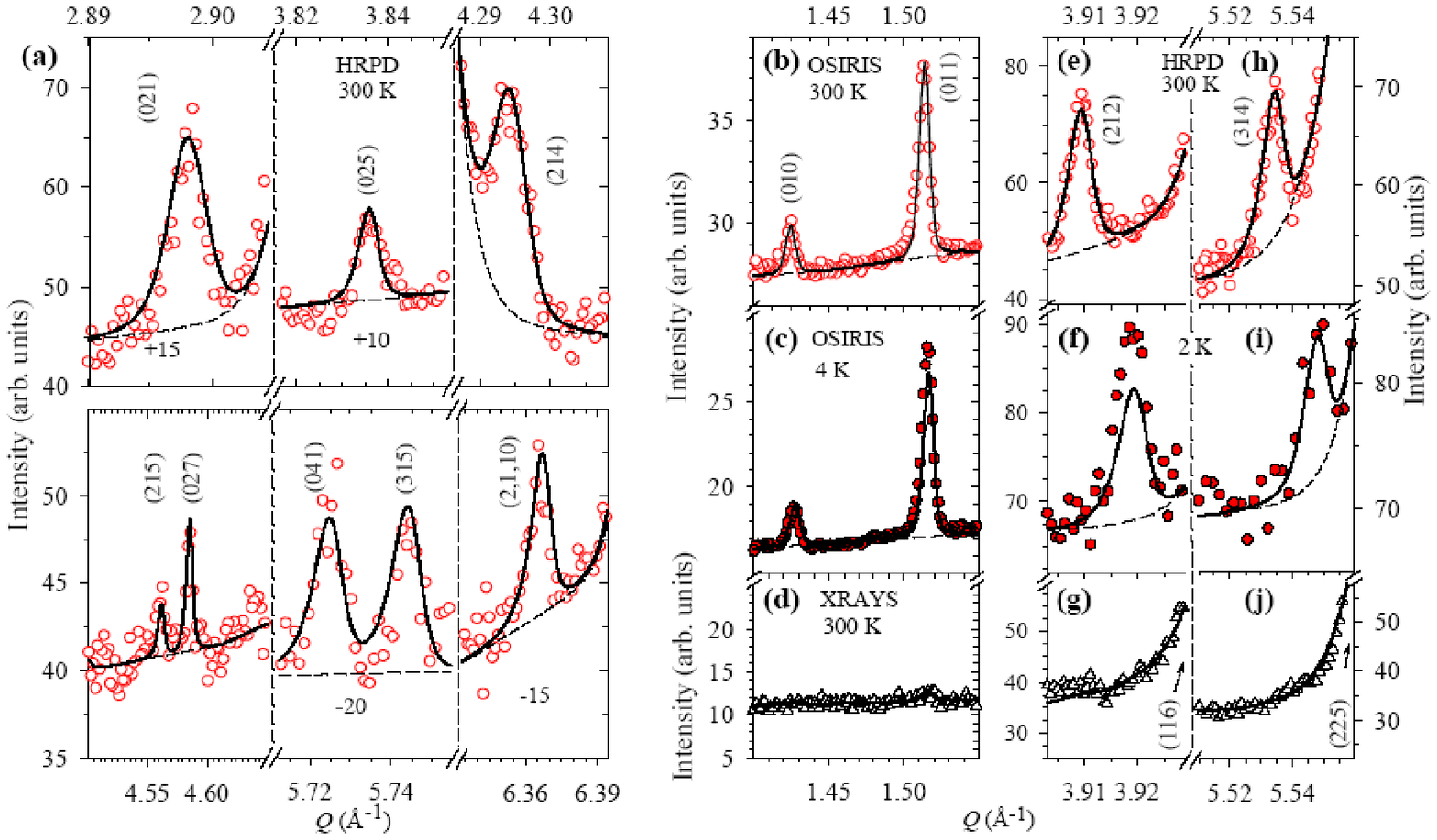}
  \caption{\label{fig_hrpd_zoom} Zoomed-in regions of the 300~K
  neutron diffraction pattern a),b),e),h) showing a number of the
  weak supercell reflections disallowed in the ideal P$6_3/mmc$
  structure and associated with a tripling of the unit cell in the
  ab plane. Solid lines are the calculated profile for the
  distorted P6$_3 2 2$ structure in Fig.\ \ref{fig_structure} and
  dashed lines show the estimated local background including the
  resolution tails of nearby main structural peaks (data in some
  subpanels is shifted vertically by indicated amounts for
  clarity). Paired panels b-c),e-f), h-i) show how the supercell
  peaks are displaced in $Q$ following the lattice contraction
  upon cooling (the base temperature data has a higher overall
  background as it was collected in a cryostat). Supercell peaks
  are not observed in the X-ray data in bottom panels d,g,j),
  consistent with the structural modulation involving mainly
  displacements of the light oxygen ions.}
\end{center}
\end{figure*}

The neutron powder diffraction pattern collected at room
temperature ($T$=300~K, Fig.\ \ref{fig_hrpd}) is overall in good
agreement with the hexagonal space group P$6_3/mmc$ proposed
before.\cite{soergel} However, a close inspection shows the
presence of a number of additional low-intensity peaks (below
$1$\% of the main peak) that could be indexed in this space group
by {\em fractional} wavevectors such as (2/3,-1/3,0) and
(2/3,-1/3,1) in Fig.\ \ref{fig_hrpd_zoom}(b), and such supercell
peaks systematically accompany the main structural peaks
throughout the wide $Q$-range probed [see panel a)] and are
displaced in $Q$ following the lattice contraction upon cooling
[see paired up-down panels b-c), e-f) and h-i)]. These extra peaks
are naturally interpreted in terms of a structural modulation
equivalent to a tripling of the unit cell in the basal plane.
Complementary X-ray measurements [Fig.\ \ref{fig_hrpd_zoom} d,g,j]
did not show a measurable intensity at the supercell positions,
suggesting that the structural modulation involves mainly
displacements of the light oxygen ions which have a very small
X-ray cross-section, as any significant displacements of the
heavier Ag or Ni ions would have implied occurrence of supercell
peaks not observed in the collected X-ray pattern. Therefore to
model the distortion we refined only the oxygen positions and
considered in order of decreasing symmetry all subgroups of the
ideal structure (P6$_3/mmc$) compatible with a tripling of the
unit cell in the ab plane, i.e. a unit cell of size $\sqrt{3}a_0
\times \sqrt{3}a_0 \times c$ (6 Ag, 6 Ni and 12 O atoms per unit
cell). We eliminated the space groups that were not compatible
with the observed low-temperature magnetic structure [see Sec.\
\ref{sec_mag_order}] where 2 Ni ions in the unit cell are
magnetically ordered and 4 are unordered, i.e. we only considered
the space groups where the magnetically-ordered and unordered Ni
ions occupied distinct crystallographic sites. The highest
symmetry subgroup in which both the structural and magnetic data
could be described is P6$_322$ (no. 182) where all 12 oxygen atoms
are symmetry-related and compared to the ideal structure are
displaced by a small amount $\epsilon$ along one of the in-plane
triangular directions. The observed supercell reflections could be
well described by this model and the best fit to the data is shown
in Figs.\ \ref{fig_hrpd} and \ref{fig_hrpd_zoom} (solid lines,
$R_{\rm Bragg} = 5.55$\%, $R_{\rm F} = 6.49$\%) (structure factors
are listed in Table.\ \ref{tab_sf_nuclear} in the Appendix A). The
obtained lattice parameters and positions in the unit cell are
listed in Table.\ \ref{tab_unitcell}. As a further test we refined
the lattice parameters using only the main peaks or only the
supercell peaks and obtained similar values ($a = 5.09082(1)$ \AA,
$c = 12.24984(4)$ \AA\ and $a = 5.0908(2)$ \AA, $c = 12.250(2)$
\AA, respectively) corroborating the fact that the weak supercell
peaks belong to the same phase as the main peaks and are not due
to an extra phase.

\begin{figure}[tbhp]
\begin{center}
  \includegraphics[width=8cm,bbllx=54,bblly=72,bburx=541,
  bbury=769,angle=0,clip=]{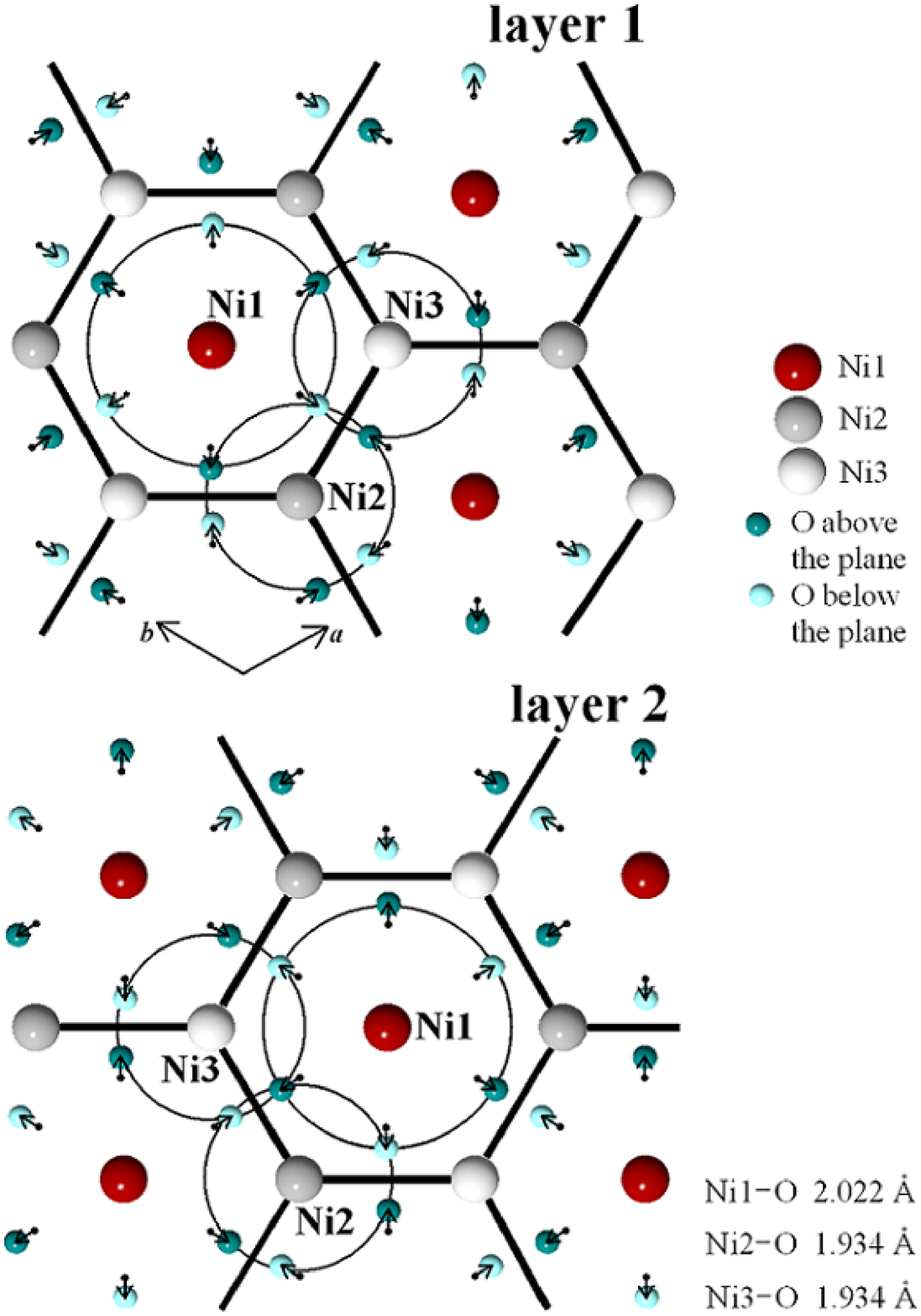}
\caption{\label{fig_structure} (color online) (top) Schematic
diagram of the NiO$_2$ layer at $z=\frac{1}{4}$ showing how the
displacements (small arrows) of the oxygen ions (small balls) lead
to a periodic arrangement of expanded (large circle, Ni1) and
contracted (small circles, Ni2,3) NiO$_6$ octahedra. Thick
hexagonal contour shows the honeycomb network of contracted sites.
The origin of the coordinate system is at the circled Ni2 site.
(bottom) The expanded site Ni1 has a staggered zig-zag arrangement
between even and odd layers stacked along the $c$-axis. Layer 2 in
the unit cell ($z=\frac{3}{4}$ and $-\frac{1}{4}$) is obtained
from layer 1 by 180$^{\circ}$ rotation around the central
($\frac{1}{2},\frac{1}{2},z$) axis followed by a ${\bm c}/2$
translation.}
\end{center}
\end{figure}

Fig.\ \ref{fig_structure} shows a schematic diagram of the $z=1/4$
NiO$_2$ layer of the distorted structure. The distortion preserves
the 3-fold rotation axis at each Ni site but the displacements of
the oxygens lead to one Ni site with expanded Ni-O bonds (Ni1) and
two other sites with contracted bonds (Ni2 and Ni3); the black
circles centered on the Ni sites correspond to Ni-O distances of
2.022 \AA\ (Ni1) and 1.934 (Ni2 and Ni3), respectively. In
Ni$^{2+}$ oxides\cite{naka} typical Ni-O bond distances are about
$\sim$ 2.09 \AA, whereas in Ni$^{4+}$ oxides\cite{tarascon} they
decrease to 1.92 \AA. This comparison suggests a charge
disproportionation among the Ni sites on the triangular layers in
2H-AgNiO$_2$ between electron-rich Ni1 sites (expanded NiO$_6$
octahedron) with valence close to Ni$^{2+}$ and electron-depleted
sites Ni2 and Ni3 (contracted octahedra) close to Ni$^{3.5+}$ (to
ensure charge neutrality). Using a phenomenological bond-valence
model\cite{bond:valence} to relate the valence of the central ion
to the bond-lengths, $v=\sum_i e^{ \left(r_0-r_i\right)/B}$ where
$B$=0.37 \AA\ and $r_0=1.686$ \AA\ for the Ni$^{3+}$-O$^{2-}$
pair, gives nominal valences in the ionic limit for the three
sites as 2.42 (Ni1) and 3.07 (Ni2 and Ni3), suggesting a
significant, but most likely only partial charge
disproportionation. We note that the bond lengths found in
2H-AgNiO$_2$ are similar to those found in YNiO$_3$
[$d$(Ni1-O)=1.923 \AA\ and $d$(Ni2-O))=1.994 \AA, ref.
\onlinecite{alonso}], proposed to have a charge disproportionation
into two Ni sublattices of valences Ni$^{3\pm \sigma}$ ($\sigma
\simeq 0.35$).

Upon cooling to lower temperatures no evidence for a further
structural distortion could be found. At 2~K the main and
supercell peaks are displaced in wavevector following the lattice
contraction and the structural peaks could be well described [see
solid lines in Fig.\ \ref{fig_hrpd_zoom} c,f,i)] by the same
crystal structure as at 300~K but with shorter lattice parameters
$a = 5.08110(2)$ \AA\ and $c = 12.24670(7)$ \AA\ ($R_{\rm Bragg} =
4.45$\%, $R_{\rm F} = 5.53$\%). We note that the related
triangular-lattice material NaNiO$_2$ behaves very differently
showing a strong ferro-distortive transition into a
low-temperature monoclinic structure with a significant difference
(4\%) in the two in-plane lattice parameters.\cite{nanio2} We
tested for such a scenario in the 2H-AgNiO$_2$ where a departure
from hexagonal symmetry could be accommodated within the
orthorhombic C$mcm$ space group with lattice parameters ${\mathrm
a} \times {\mathrm b} \times {\mathrm c}$. Within the experimental
accuracy no splitting of the main peaks could be detected [see
Fig.\ \ref{fig_hrpd}b)] and the fitted in-plane lattice parameters
had the same ratio as in the undistorted structure (${\mathrm
b}/{\mathrm a}=\sqrt{3}$) to within better than 0.02\%, so we
concluded that the hexagonal symmetry is preserved down to the
lowest temperature probed of 2~K.

\begin{figure*}[tb]
\begin{center}
  \includegraphics[width=18cm,bbllx=54,bblly=323,bburx=540,
  bbury=522,angle=0,clip=] {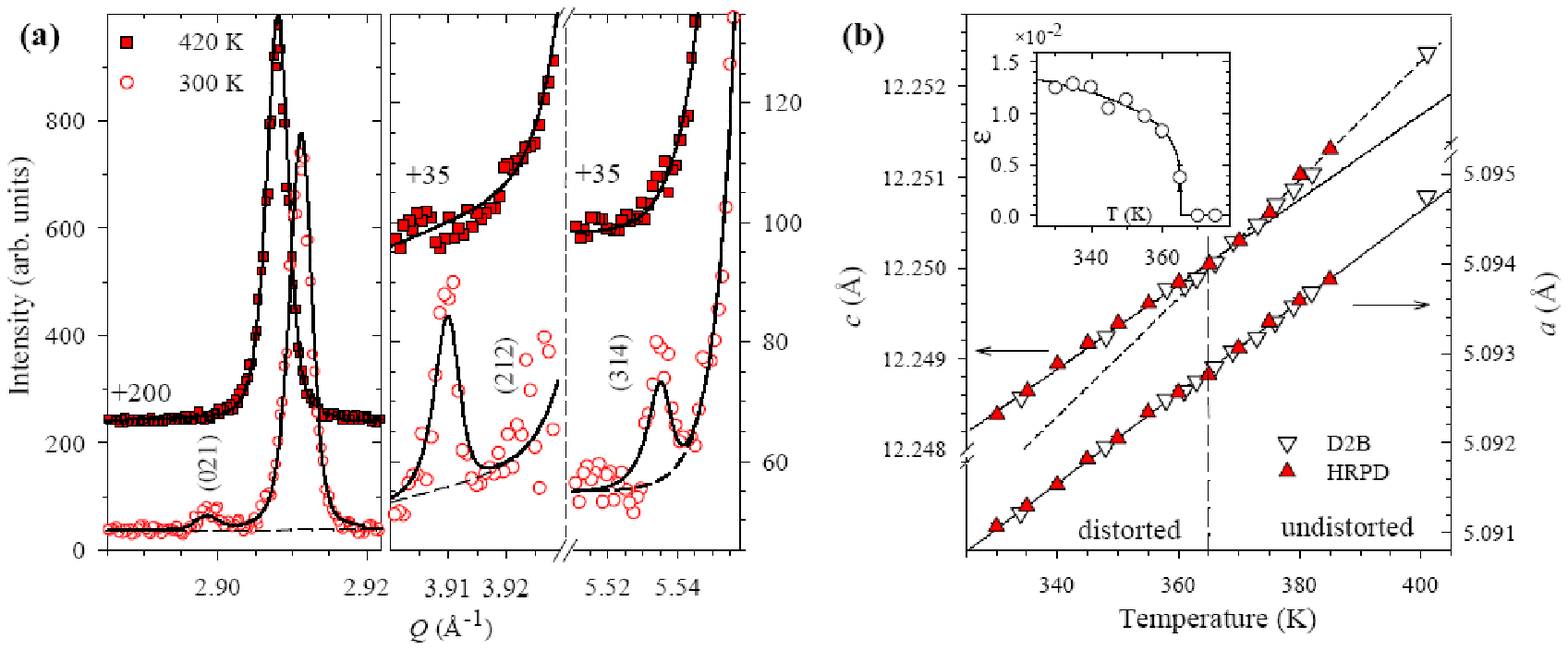}
  \caption{\label{fig_hrpd_hight} (a) Comparison between 300~K
  (open symbols, lower trace) and 420~K data (filled symbols,
  upper trace) showing the absence of the triple-cell peaks (021),
  (212) and (314) at high temperature. Solid lines are fits to the
  distorted (300~K) and ideal structure (420~K), respectively and
  dashed lines show the estimated background level. The
  high-temperature data has been shifted vertically by the
  indicated offsets for clarity. (b) Temperature dependence of the
  lattice parameters: $c$ left axis, $a$ right axis. Solid (dashed)
  lines for the $c$ parameter are straight line fits to the data
  below (above) $T_S=365$~K. Filled/open symbols are data from
  different instruments. Inset: temperature-dependence of the
  oxygen displacement parameter $\epsilon$, solid line is guide to
  the eye.}
\end{center}
\end{figure*}

\section{Transition to the high-symmetry structure at high temperatures}
\label{sec_hight}
We also measured the diffraction pattern at high temperatures
motivated by recent resistivity measurements\cite{soergel07}
reporting a weak anomaly near 365 K, proposed to originate from a
structural transition. We observed that upon heating the
triple-cell peaks decreased in intensity and could not be observed
above $T_S=365(3)$~K, which coincides with the location of the
transport anomaly. Fig.\ \ref{fig_hrpd_hight}(a) shows that
triple-cell peaks are absent in the 420 K data and the main
structural peaks are only slightly displaced in $Q$ due to lattice
expansion. The collected diffraction patterns at various
temperatures were refined in the distorted space group P6$_3$22
and the obtained temperature-dependence of the oxygen displacement
$\epsilon$ away from the high-symmetry position is plotted in
Fig.\ \ref{fig_hrpd_hight}(b) inset and shows a rather rapid
decrease near $T_S$. The lattice parameters increase smoothly with
increasing temperature with no clear anomalies near the transition
apart from possibly a small kink in the $c$ lattice constant as
linear fits to the data below and above 365~K (solid and dashed
lines in Fig.\ \ref{fig_hrpd_hight}(b)) give a slightly smaller
slope in the low-temperature phase, consistent with the structural
distortion in this phase making the crystal lattice more rigid so
less able to expand upon increasing temperature. The disappearance
of the supercell peaks at high temperature indicates a transition
to the high-symmetry crystal structure where all Ni-O bonds become
equivalent and charge is uniformly distributed on all Ni sites.
Indeed the 420~K data could be well described by the undistorted
P6$_3/mmc$ space group with $a_0=2.94267(3)$ \AA, $c=12.2554(2)$
\AA\ and oxygen height $z_{\rm O}=0.07991(6)$ ($R_{\rm
Bragg}=11.8$ \%, $R_{\rm F}=7.13$ \%).

To conclude the analysis of the diffraction pattern we note that
in addition to the triple cell peaks identified above, the data
also showed diffraction peaks due to a small admixture (1\%) of
the rhombohedral 3R polytype (currently at the limit at which pure
hexagonal 2H polytype can be chemically prepared) and some other
small peaks that could not be indexed by any obvious commensurate
fractional index of the main peaks and which were still present at
high temperatures above the structural transition at 365 K. Those
were attributed to unidentified impurity phases below the 1\%
level that occurred during chemical synthesis (scanning electron
microscopy (SEM) coupled with energy-dispersive X-ray analysis
(EDX) of a small part of the sample indicated small traces of Au,
Al and Si, but the chemical composition could not be precisely
identified). The alternative origin of those small peaks could
have been instrumental (e.g. spurious sample holder reflections).

\begin{table}[tb]
\begin{center}
\caption{\label{tab_co} Basis vectors for the irreducible
representations for charge (scalar) order at the Ni sites (2$a$)
in space group P$6_3/mmc$ (no. 194) for propagation vector ${\bm
q}_0=(1/3, 1/3, 0)$ obtained using group theory\cite{sikora}.}
\begin{tabular}{l|c|c} \hline
position & (0, 0, 0) & (0, 0, $\frac{1}{2}$) \\[2pt]
\hline
$\Gamma_1$ & 1 & 1 \\
$\Gamma_2$ & 1 & -1 \\
\hline
\end{tabular}
\end{center}
\end{table}

\begin{table*}[tb]
\begin{center}
\caption{\label{tab_po} Basis vectors of the irreducible
representations for displacement (polar vector) order at the
oxygen sites (4$f$) in space group P$6_3/mmc$ (no. 194) for
propagation vector ${\bm q}_0=(1/3, 1/3, 0)$ obtained using group
theory\cite{sikora}.}
\begin{tabular}{l|c|c|c|c} \hline
position & ($\frac{2}{3}$, $\frac{1}{3}$, $z_O$) & ($\frac{1}{3}$,
$\frac{2}{3}$, 1-$z_O$) & ($\frac{2}{3}$, $\frac{1}{3}$,
$\frac{1}{2}-z_O$) & ($\frac{1}{3}$, $\frac{2}{3}$,
$\frac{1}{2}+z_O$)
\\[2pt]
\hline $\Gamma_1$ & ($e^{i\pi/6}, e^{i\pi/2}, 0$) & ($e^{i\pi/2},
e^{i\pi/6},0$) & ($e^{i\pi/6}, e^{i\pi/2}, 0$) & ($e^{i\pi/2},
e^{i\pi/6},0$) \\[2pt]
$\Gamma_2$ & ($e^{i\pi/6}, e^{i\pi/2}, 0$) & ($e^{i\pi/2},
e^{i\pi/6},0$) & $-(e^{i\pi/6}, e^{i\pi/2}, 0)$ & $-(e^{i\pi/2},
e^{i\pi/6},0)$ \\[2pt]
\hline
\end{tabular}
\end{center}
\end{table*}

\section{Symmetry analysis of the structural distortion}
\label{sec_sym}
To understand better the mechanism of the structural transition at
$T_S=365$ K from the ideal to the distorted structure we performed
an analysis of symmetry-allowed order patterns in the undistorted,
high-temperature space group P$6_3/mmc$ with the observed ordering
wavevector ${\bm q}_0=(1/3,1/3,0)$. We considered both the charge
(scalar) order on the Ni sites as well as the oxygen ion
displacements (polar vector order).

Our refined model for the structure in the distorted phase in
Fig.\ \ref{fig_structure} shows a periodic arrangement of expanded
(Ni1) Ni$^{3-\sigma}$ and contracted (Ni2 and Ni3)
Ni$^{3+0.5\sigma}$ sites with $\sigma=1$ in case of complete
charge disproportionation. To try to reproduce this we looked for
symmetry-allowed charge (scalar) order patterns with propagation
vector ${\bm q}_0=(1/3,1/3,0)$ at the Ni $2a$ sites (2 atoms) in
the unit cell of the P$6_3/mmc$ space group. The obtained
irreducible representations and basis vectors are listed in Table\
\ref{tab_co}. There are two basis vectors ($\tau_{1,2}$
corresponding to irreducible representations $\Gamma_{1,2}$ in
Table \ref{tab_co}) which physically correspond to having the same
or opposite charges for Ni sites above each other in the two
layers. The order pattern implied by the structural refinement
cannot be described by a single basis vector, but can be described
by a linear combination of two basis vectors, i.e.
\begin{equation}
\label{eq_zeta} \zeta=e^{i\pi/3}\left[ \left(\tau_1
+\tau_{2}\right) + e^{i4\pi/3} \left(\tau_1
-\tau_{2}\right)\right],
\end{equation}
where $\tau_{i}(1)$ is the charge on atom 1 (layer 1) for the
basis vector $\tau_{i}$ etc.

\begin{figure}[tbhp]
\begin{center}
 \includegraphics[width=7.8cm,bbllx=54,bblly=298,bburx=542,
  bbury=545,angle=0,clip=]{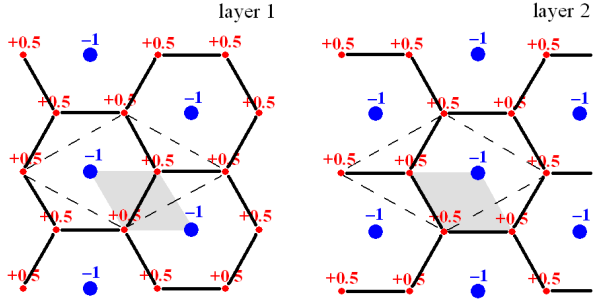}
  \caption{\label{fig_co} (color online) Charge order pattern
  described by eq.\ (\ref{eq_co}) to be compared with Fig.\
  \ref{fig_structure}. Labels -1 and +0.5 indicate charges in
  units of $\sigma$. Thick hexagonal contours indicate the honeycomb
  network of the electron depleted Ni2,3 sites and dashed line
  contour is the unit cell of the charge-ordered structure. Light
  shaded area is the unit cell of the ideal structure with
  all Ni sites equivalent.}
\end{center}
\end{figure}

The corresponding charge order pattern (relative to the uniform
high-temperature phase where each Ni site has valence $+3$) is
shown in Fig.\ \ref{fig_co} and consists of a triangular lattice
of charges $-\sigma$, i.e. electron-rich Ni$^{3-\sigma}$,
surrounded by a honeycombe network of charges $+0.5\sigma$, i.e.
Ni$^{3+0.5\sigma}$. This charge order pattern is obtained from the
basis vector $\zeta$ modulated by the phase factor due to the
propagation vector $\bm{q_0}$, i.e. in the unit cell at distance
$\bm{r}=n_1\bm{a}_0 +n_2\bm{b}_0+n_3\bm{c}$ from the origin
($n_{1,2,3}$ integers) the charges are given by the real part of
the complex vector
\begin{equation} \tilde{Q}_{\bm{r}}= \frac{\sigma}{2} e^{i\bm{q}_0
\bm{r}}\zeta.
\label{eq_co}
\end{equation}
For atom 1 (layer 1) the charge is
\begin{equation}
Q_{\bm{r}}(1)= \sigma \cos \left( \bm{q}_0 \bm{r}+ \pi/3\right)
\end{equation}
and for atom 2 (layer 2) is
\begin{equation}
Q_{\bm{r}}(2)= \sigma \cos \left( \bm{q}_0 \bm{r}+ 5 \pi/3\right).
\end{equation}
The normalization prefactor in eq.\ (\ref{eq_co}) gives the
magnitude of the charge order and was chosen such as to obtain
charge $-\sigma$ on the Ni1 sites.

Subsequently we looked at possible order patterns for oxygen ion
displacements corresponding to the same propagation wavevector
${\bm q}_0$. There are in total 4 one-dimensional and 2
two-dimensional irreducible representations that could describe
displacement (polar vector) order at the 4$f$ oxygen sites (4
atoms per unit cell) and the two one-dimensional representations
relevant for our discussion are listed in Table\ \ref{tab_po}.
Basis vectors of these representations transform in the same way
under the symmetry operations of the P$6_3/mmc$ space group as the
basis vectors for charge order listed in Table\ \ref{tab_co} so we
use the same symbols $\tau_1$ and $\tau_2$. Here they have complex
components indicating a degeneracy with respect to rotation of the
displacement vector in the $ab$ plane. We again find that the
observed structural displacements cannot be described by a single
basis vector, but is described by a linear combination of two
basis vectors, i.e.
\begin{equation}
\label{xi} \bm{\xi}=\frac{1}{2}e^{i7\pi/6}\left[ \left(\bm{\tau}_1
+\bm{\tau}_{2}\right) + e^{i4\pi/3} \left(\bm{\tau}_1
-\bm{\tau}_{2}\right)\right], \label{eq_po}
\end{equation}
where the complex phase factor in front serves to rotate the
oxygen displacement vector in the $ab$ plane and with the chosen
phase the displacement pattern is as in Fig.\ \ref{fig_structure},
i.e. oxygens displaced radially out of the expanded Ni sites (the
numerical prefactor is used for normalization). In the above
equation $\bm{\tau}_i(1)$ is the complex displacement vector for
atom 1 in the representation $\bm{\tau}_i$ as per Table\
\ref{tab_po}. (Note that we use bold symbols to indicate vectors
with components along the three crystallographic directions). The
relative phase factor of the $\bm{\tau}_1$ and $\bm{\tau}_2$ basis
vectors is the same for the polar order of oxygen displacements
eq.\ (\ref{eq_po}) and for the charge order on the Ni sites eq.\
(\ref{eq_zeta}) as expected if the structural transition was
driven by charge fluctuations coupled to an oxygen phonon
breathing mode. It is noteworthy that the obtained oxygen
displacements are consistent with an optic phonon mode at the
Brillouin zone corner point $\bm{q}_0$ predicted by lattice
dynamics calculations\cite{bejas} for a structure identical to one
undistorted NiO$_2$ layer. So our results provide support that it
is this phonon mode (probably modified slightly due to couplings
with the Ag layer above and below) that mediates the $\sqrt{3}
\times \sqrt{3}$ charge order in the triangular NiO$_2$ planes.

In appendix B we provide further symmetry analysis of the basis
vector modes for the charge and displacement orders using
co-representation analysis and prove that they uniquely identify
the distorted space group as P6$_3$22.

To conclude this section for completeness we quote below the
expressions for obtaining the individual atomic displacements
starting from the basis vector mode $\bm{\xi}$. The order pattern
is given by $\bm{\xi}$ modulated by the phase factor due to the
propagation vector $\bm{q}_0$, i.e.
\begin{equation}
\bm{\tilde{d}}_{\bm{r}}= 2\epsilon ~e^{i\bm{q}_0 \bm{r}}~\bm{\xi},
\end{equation}
where $\bm{r}=(n_1,n_2,n_3)$ defines the unit cell and
$\bm{\xi}(1)$ is the basis vector for atom 1 etc. The scale
prefactor $2\epsilon $ was introduced such as to give the absolute
magnitude of the final displacements $\epsilon\sqrt{3}a_0$ as per
Table\ \ref{tab_unitcell}. The real part of the above equation
gives the actual displacement vectors
\begin{equation}
2\epsilon \left( \cos \left( \bm{q}_0 \bm{r}+ \phi_x \right),\cos
\left( \bm{q}_0 \bm{r}+ \phi_y \right),0 \right)
\end{equation}
where the phases $\phi_{x,y}$ are given by the basis vector for
that atom $\bm{\xi}= (e^{i\phi_x},e^{i\phi_y},0)$ using eq.\
(\ref{eq_po}) and Table\ \ref{tab_po}. For example for the oxygen
atom 1 in the unit cell at the origin $\bm{r}=(0,0,0)$ the
displacement is
\begin{equation}
2\epsilon \left( \cos \left( 4\pi/3 \right),\cos \left( 5\pi/3
\right),0 \right)=\epsilon (-1,1,0).
\end{equation}

\section{Susceptibility and specific heat}
\label{sec_sus}

Magnetic susceptibility measured on a small powder sample is shown
in Fig.\ \ref{fig_chi}(a). The pronounced drop at low temperatures
near 20 K [see inset] is attributed to the onset of magnetic order
and at higher temperatures above $\sim70$ K the data can be well
described by a local-moment Curie-Weiss form plus a small
temperature-independent part $\chi_0$,
\begin{equation}
{\cal \chi}=\frac{C}{T-\theta}+\chi_0 \label{eq_chi}
\end{equation}
where $C=0.445(5)$ emu K/mole, $\theta_{\rm {CW}}=-107(2)$ K, and
$\chi_0=1.7(1) \times 10^{-4}$ emu/mole, similar to previous
reports.\cite{soergel} The large negative Curie-Weiss temperature
shows dominant antiferromagnetic interactions. Magnetic order
occurs only at significantly lower temperatures, $\sim$20~K,
suggesting that fluctuations due the low-dimensionality (mainly
in-plane interactions) and the frustrated triangular geometry are
important in suppressing the magnetic ordering temperature.

The effective magnetic moment extracted from the Curie-Weiss fit is
$\mu_{\rm eff}=1.88$ $\mu_B$ per Ni ion, and this was used
previously \cite{soergel} as evidence that Ni sites were in the
low-spin Ni$^{3+}$ state ($t_{2g}^6e_g^1$) with $S=1/2$ ($\mu_{\rm
eff}=1.73 \mu_B$ for $g=2$). However, a charge disproportionation
scenario as suggested by our structural measurements could also lead
to similar values of the effective moment. In the extreme case of
complete charge disproportionation the Ni1 site is Ni$^{2+}$
$t_{2g}^6e_g^2$ with a large spin moment $S=1$, whereas the other
sites are Ni$^{3.5+}$ likely to have only a very small spin moment
as they are close to Ni$^{4+}$ with a filled $t_{2g}^6$ level, which
has $S=0$. The average effective moment observed by high-temperature
susceptibility considering only $S=1$ moments on the Ni1 sites
would then be $\mu_{\rm eff}=1.63 \mu_B$ (for $g=2$), which is only
$6$\% lower than the value for $S=1/2$ at every site, and it is
possible that including a more realistic scenario of partial charge
disproportionation (more likely to be the case here) could bring
this estimate closer to the experiment.

Finally, we plot in Fig.\ \ref{fig_chi}(b) specific heat
measurements which observe a sharp lambda-like peak identified
with the magnetic transition near 20~K. No other anomalies could
be observed up to the highest temperature studied of 240~K,
consistent with no additional (structural) transitions occurring
in this temperature range.

\section{Magnetic structure}
\label{sec_mag_order}
\begin{figure}[tbhp]
\begin{center}
  \includegraphics[width=7.8cm,bbllx=63,bblly=258,bburx=446,
  bbury=512,angle=0,clip=] {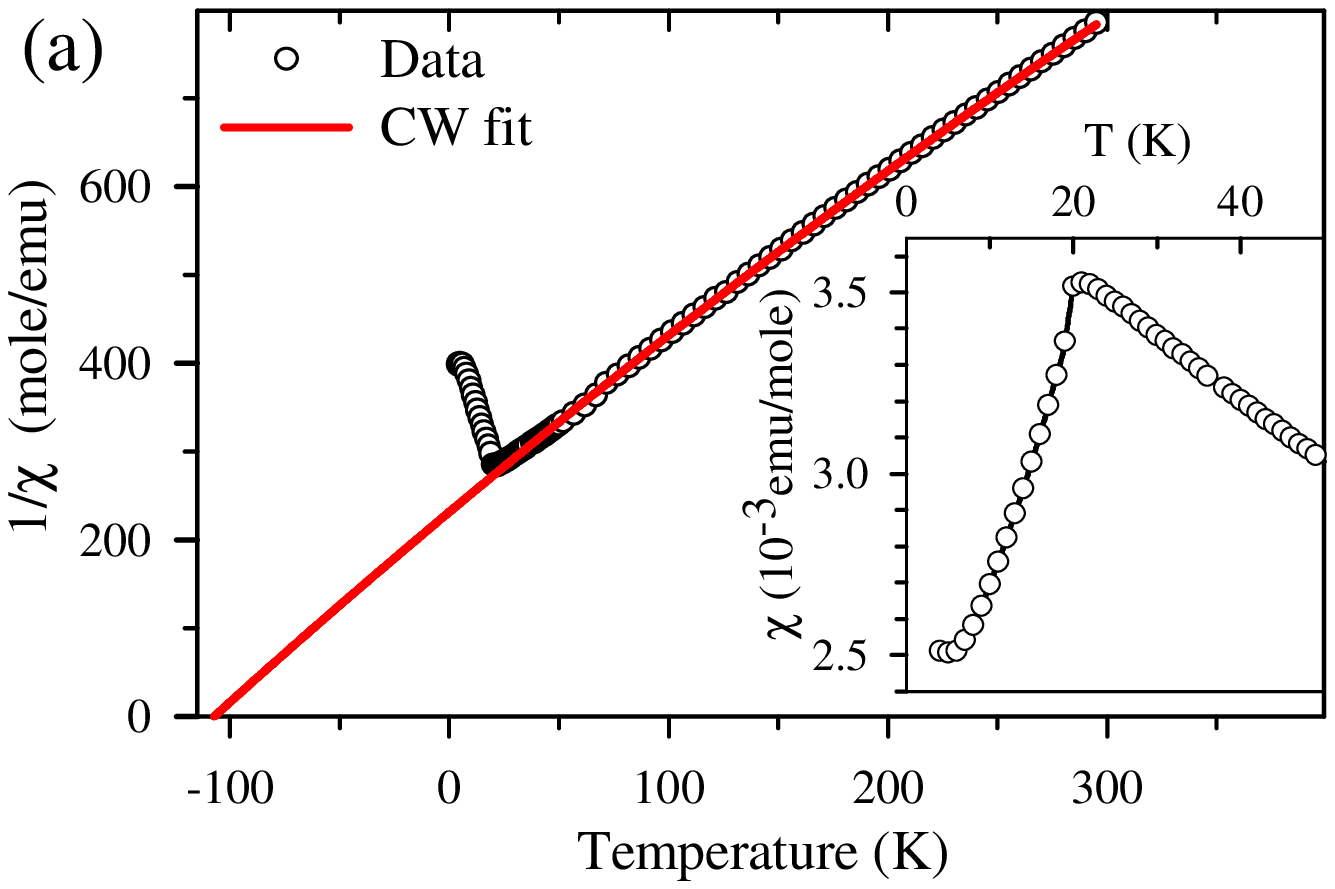}
  \vspace{0.4cm}
  \includegraphics[width=7.5cm,bbllx=73,bblly=309,bburx=442,
  bbury=563,angle=0,clip=] {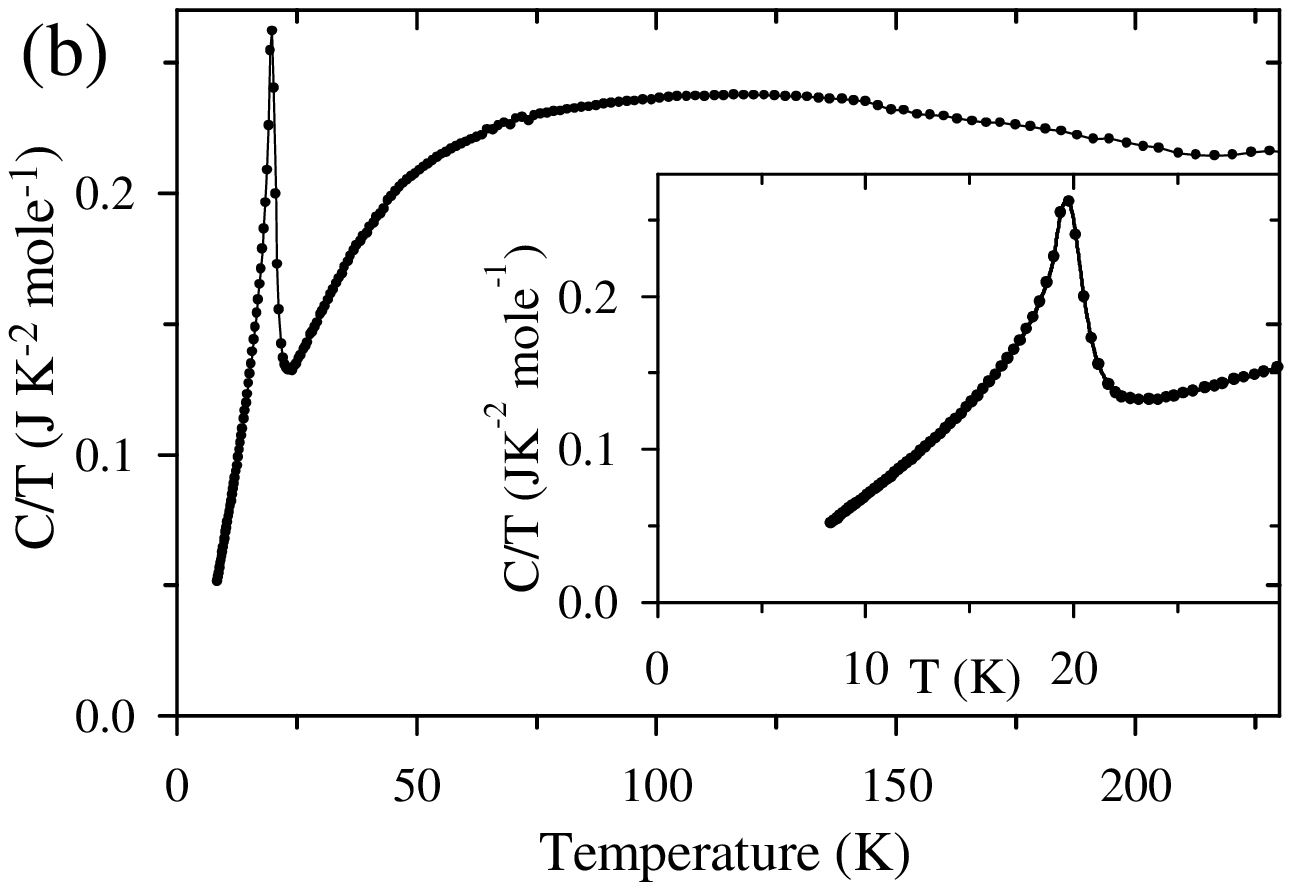}
  \caption{\label{fig_chi}(a) Inverse magnetic susceptibility ($1/\chi$)
  fitted to a Curie-Weiss law eq.~(1) (solid line) gives a large negative
  intercept indicating dominant antiferromagnetic interactions.
  Inset shows suppression of susceptibility below 20 K attributed
  to onset of antiferromagnetic order. (b) Temperature dependence
  of the specific heat observing a sharp lambda-like peak near
  the magnetic transition temperature.}
\end{center}
\end{figure}

\begin{figure}[tbhp]
\begin{center}
  \includegraphics[width=8cm,bbllx=54,bblly=256,bburx=537,
  bbury=588,angle=0,clip=] {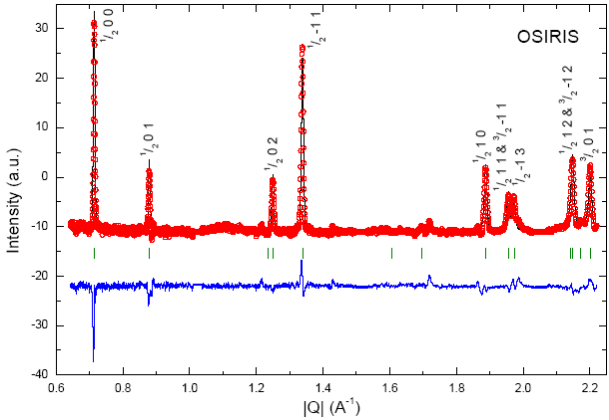}
  \caption{\label{fig_mag_pattern}
Difference pattern 4~K $-$ 300~K from OSIRIS showing peaks of
magnetic origin, indexed by the propagation vector ${\bm
k}$=(1/2,0,0). The circles represent the observed intensities, the
solid curve is a fit to the magnetic structure depicted in Fig.\
\ref{fig_mag_struct} and vertical bars indicate the magnetic Bragg
peak positions. The bottom curve shows the difference between the
fit and the data.}
\end{center}
\end{figure}

\begin{figure}[tbhp]
\begin{center}
 \includegraphics[width=7.5cm,bbllx=93,bblly=230,bburx=500,
    bbury=537,angle=0,clip=] {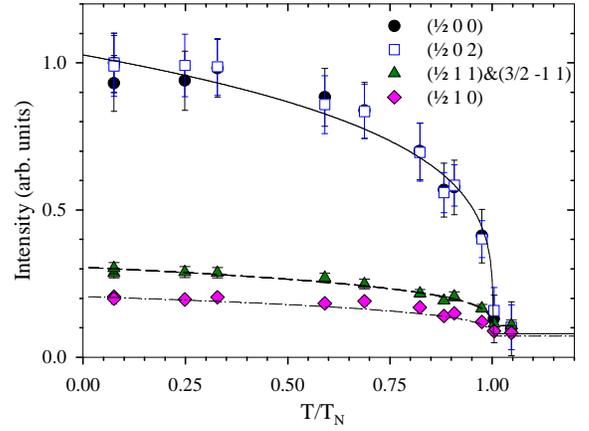}
  \caption{\label{fig_mag_order_par}
Observed intensity of several magnetic Bragg peaks as a function
of reduced temperature $T/T_N$. Data points are from elastic
neutron powder scattering measurements using IN6. Lines are guides
to the eye.}
\end{center}
\end{figure}

The magnetic susceptibility shown in Fig.\ \ref{fig_chi} (a) inset
has a sharp downturn below 20\ K as characteristic of a transition
to antiferromagnetic order and below this temperature additional
reflections are observed in the neutron diffraction pattern at low
$Q$. The magnetic order peaks are most easily seen in the
difference pattern 4~K $-$ 300~K shown in Fig.\
\ref{fig_mag_pattern} and can be indexed with respect to the P6$_3
2 2$ supercell by the commensurate propagation vector ${\bm
k}=(1/2,0,0)$. The magnetic order parameter curves are plotted in
Fig.\ \ref{fig_mag_order_par} as a function of reduced temperature
$T/T_{\rm N}$. The N\'{e}el temperature obtained from the neutron
data is 23.7(3)~K is slightly higher than the location of the
specific heat anomaly near 19.7(3)~K, and susceptibility maximum
near 21(1)~K. This is probably due to a small temperature
calibration offset in the neutron measurement where the sensor was
at some distance away from the sample. We regard the absolute
value of the transition temperature observed by the specific heat
measurement $T_N$=19.7(3)~K as the most accurate, in agreement
with recent $\mu$SR results.\cite{musr}

To find the magnetic structure we first consider in Table\
\ref{tab_group_theory} the magnetic basis vectors compatible with
the symmetry of the P$6_3 2 2$ crystal structure for the six Ni
ions in the unit cell (three sublattices). Symmetry constrains the
moments on the same sublattice to be either parallel or
antiparallel between the two layers in the unit cell. The best fit
to the observed diffraction pattern (varying the spin direction
and magnitude independently on the three sublattices) was obtained
for the case when only one of the three sublattices was ordered,
either Ni1 or Ni3, with a moment of 1.552(7)\ $\mu_B$ along the
$c$-axis and spins parallel between adjacent layers (irreducible
representation $\Lambda_3$ in Table\ \ref{tab_group_theory} with
$v \ne 0$ or $v'' \ne 0$ and all the other spin components equal
to 0). This correlates well with the structural analysis whereby
the Ni1 site should have a large spin moment ($S=1$ at complete
charge disproportionation Ni$^{2+}$) whereas the other two sites
(Ni2 and Ni3) with valences close to Ni$^{3.5+}$ would have a
much smaller moment, as their configuration would be close to
Ni$^{4+}$ with $S=0$. Therefore we associate the magnetic order
with site Ni1 carrying the largest spin moment.

The experimental data doesn't preclude the possibility of having
also a very small magnetic moment along the $c$-axis on either Ni2
or Ni3 sites, these however could only be of the order of 0.1
$\mu_B$ at maximum (the spin arrangement would then be described
by $\Lambda_3$ with $v'$ or $v^{\prime\prime}$ equal to $-0.1$\
$\mu_B$, respectively). Because taking into account these small
magnetic moments doesn't improve the quality of fit in the
following we assume for simplicity that only the sublattice Ni1 is
ordered with a large spin moment.

\begin{figure}[tbhp]
\begin{center}
  \includegraphics[width=8cm,bbllx=56,bblly=319,bburx=531,
  bbury=521,angle=0,clip=] {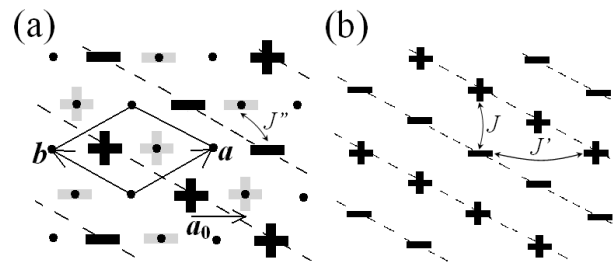}
\caption{\label{fig_mag_struct} (a) Magnetic structure of the Ni1
sublattice consisting of alternating rows (dashed lines) of
ferromagnetically aligned spins. $\pm$ symbols indicate the
projection of the spin moment along the $c$-axis, dots represent
the unordered (Ni2 and Ni3) Ni sites in the unit cell (solid
contour). Thick black symbols indicate the moments in the bottom
NiO$_2$ plane ($z=\frac{1}{4}$) whereas faint grey symbols
correspond to the moments in the upper plane ($z=3/4$) obtained by
shifting the pattern by an in-plane offset $\bm{a_0}$. The
arrowed line labelled $J^{\prime\prime}$ indicates one of the
three inter-layer exchange paths. The drawn structure has
propagation vector ${\bm k}=(1/2,0,0)$ (Bragg peaks indicated by
black stars in Fig.\ \ref{fig_rec_space}), equivalent structures
are obtained by $\pm 60^{\circ}$ rotation around the
$(\frac{2}{3},\frac{1}{3},z)$-axis. (b) In a single layer the
ordered sites form a triangular lattice of spacing $a$. Short and
long arrowed lines indicate the nearest- and
next-nearest-neighbour exchanges $J$ and $J^{\prime}$ in a minimal
model proposed to explain the stability of the observed
structure.}
\end{center}
\end{figure}

The magnetic structure is illustrated in Fig.\
\ref{fig_mag_struct}(a) and consists of alternating ferromagnetic
stripes, i.e. spins are parallel along one of the three directions
in the triangular plane and antiparallel along the other two
directions. This structure can occur in three different domains
obtained by $\pm60^{\circ}$ rotation around the $c$-axis, which
would correspond to the symmetry-equivalent wavevectors ${\bm
k^{\prime}}$=(0,1/2,0) and ${\bm k^{\prime\prime}}$=(1/2,-1/2,0)
of the star of ${\bm k}$. We note that the magnetic structure has
only a 2-fold symmetry rotation axis along the $c$-axis whereas
the crystal structure has 3-fold rotation symmetry along the
$c$-axis. Three equal-weighted domains of those three structures
would be expected in a macroscopic sample and each domain has the
same powder-averaged diffraction pattern.

In the following we discuss possible mechanisms to stabilize the
observed magnetic structure in a model of stacked triangular
layers, so we consider the Hamiltonian
\begin{equation} {\cal H}=\sum_{\rm{NN}} J {\bm S}_i \cdot
\bm{S}_j+\sum_{\rm{NNN}}J^{\prime} {\bm S}_i \cdot \bm{S}_k
+\!\!\!\!\!\sum_{\rm interlayer} \!\!\!\!\!\!J^{\prime\prime}{\bm
S}_i \cdot \bm{S}_l -D\sum_i\left(S_i^z\right)^2. \label{eq_ham}
\end{equation}
Here NN indicates summing over all in-plane nearest-neighbor pairs
with coupling $J$, NNN denotes next-nearest-neighbour in-plane pairs
with coupling $J^{\prime}$ [see Fig.\ \ref{fig_mag_struct}(b)] and
$J^{\prime\prime}$ is the inter-layer coupling (3 neighbors above
and 3 below, see Fig.\ \ref{fig_mag_struct}(a)), with
$J,J^{\prime}>0$ (antiferromagnetic) and $J^{\prime\prime}<0$
(ferromagnetic). The last term in eq.\ (\ref{eq_ham}) is an on site
easy-axis anisotropy proposed to arise from crystal-field effects
and required to stabilize the ordering spin direction along the
crystallographic $c$-axis.

Considering first the 2D antiferromagnetic Heisenberg model
($D=J^{\prime\prime}=0$ in eq.\ (\ref{eq_ham})) the ground state
for NN couplings only is the 3-sublattice 120$^{\circ}$ spiral,
however adding moderate antiferromagnetic NNN couplings $1/8
~\rlap{\lower4pt\hbox{\hskip1pt$\sim$}}
    \raise1pt\hbox{$<$}~J^{\prime}/J\le1$ stabilizes the collinear
stripe order.\cite{jolicoeur} In fact for this model at the
classical level the two-sublattice stripe order is degenerate with
a continuous manifold of 4-sublattice non-collinear states but
zero-point quantum fluctuations are predicted to lift this
degeneracy through ``order by disorder" and select the stripe
ground state. An easy axis anisotropy $D>0$ is expected to further
stabilize the collinear state thereby reducing the minimal
required $J^{\prime}$. At the classical level in the limit of very
large anisotropy approaching the Ising limit the required
$J^{\prime}$ becomes infinitesimally small, but is non
zero.\cite{slotte} In particular the NNN couplings are required to
lift the degeneracy between the stripe state and the collinear
state with 2 spins up and one down for each triangle, the
so-called up-up-down state (UUD), with the ground state energy per
spin $e^{\rm S}=(-J-J^{\prime})S^2$ lower than $e^{\rm
UUD}=(-J+3J^{\prime})S^2$ for antiferromagnetic $J^{\prime}$. As
an alternative to NNN couplings we note that the interlayer
interactions could also provide a mechanism to lift this
degeneracy as the stacking of magnetic layers energetically favors
the stripe order. The interlayer energy is $J^{\prime\prime}S^2$
(energy gain for $J^{\prime\prime}<0$ ferromagnetic) for the
stacked stripe order depicted in Fig.\ \ref{fig_mag_struct}(a)
where each spin has 4 favorable and 2 unfavorable interlayer
bonds, whereas for a stacked UUD structure the interlayer energy
is reduced to $J^{\prime\prime}S^2/3$ because only $2/3^{\rm rds}$
of sites have 2 net favorable bonds and the remaining $1/3^{\rm
rd}$ have 2 net unfavorable interlayer bonds. From this we
conclude that the observed magnetic structure could be explained
starting from an easy-axis nearest-neighbor triangular
antiferromagnet with additional weak in-plane second neighbor
antiferromagnetic exchange or weak ferromagnetic interlayer
couplings.

\begin{figure}[tbhp]
\begin{center}
  \includegraphics[width=7.0cm,bbllx=53,bblly=194,bburx=540,
  bbury=648,angle=0,clip=] {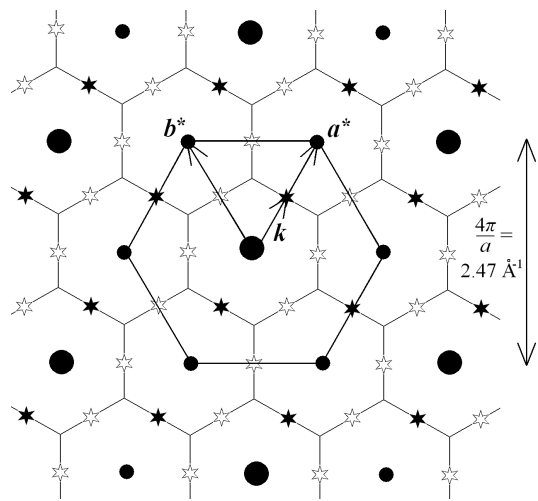}
  \caption{\label{fig_rec_space} Reciprocal basal plane showing
  locations of structural (filled circles) and magnetic
  reflections (stars). Large filled circles and large bold hexagon
  indicate the zone centers and Brillouin zone of the ideal,
  undistorted structure (P6$_3$/$mmc$), whereas small filled
  circles and thin line hexagons are respectively structural
  supercell peaks and the Brillouin zone of the distorted crystal
  structure (P6$_3 2 2$). Filled stars are the magnetic Bragg peak
  positions from the magnetic structure with propagation
  wavevector ${\bm k}=(1/2,0,0)$ shown in Fig.\
  \ref{fig_mag_struct}, open stars are peaks from domains rotated
  by $\pm 60^{\circ}$.}
\end{center}
\end{figure}

\begin{table*}[tb]
\begin{center}
\caption{\label{tab_group_theory} Basis vectors for irreducible
representations for magnetic order at the six Ni ions in the unit
cell (3 independent sublattices) for a structure with propagation
vector ${\bm k}$=(1/2,0,0) for the P6$_3 2 2$ space group obtained
using group theory (MODY package\cite{sikora}). $u$, $v$, $u'$,
$v'$, $u''$ and $v''$ are independent spin components.}
\begin{tabular}{|l|c|c|c}
\hline
\begin{tabular}{l}
site\\
\hline
position\\[2pt]
\hline
$\Lambda_1$\\
$\Lambda_2$\\
$\Lambda_3$\\
$\Lambda_4$\\
\hline
\end{tabular}
&
\begin{tabular}{c|c|c}
Ni1 (2$c$) & Ni2 (2$b$) & Ni3 (2$d$)\\
\hline
\begin{tabular}{c|c}
($\frac{1}{3}$, $\frac{2}{3}$, $\frac{1}{4}$) & ($\frac{2}{3}$, $\frac{1}{3}$,
$\frac{3}{4}$)\\[2pt]
\hline
($2u$,$u$,0) & ($2u$,$u$,0)\\
(0,-$u$,$v$) & (0,-$u$,-$v$)\\
(0,-$u$,$v$) & (0,$u$,$v$)\\
($2u$,$u$,0) & (-$2u$,-$u$,0)\\
\end{tabular}
&
\begin{tabular}{c|c}
(0, 0, $\frac{1}{4}$) & (0, 0, $\frac{3}{4}$) \\[2pt]
\hline
($2u'$,$u'$,0) & (-$2u'$,-$u'$,0)\\
(0,-$u'$,$v'$) & (0,$u'$,$v'$)\\
(0,-$u'$,$v'$) & (0,-$u'$,-$v'$)\\
($2u'$,$u'$,0) & ($2u'$,$u'$,0)\\
\end{tabular}
&
\begin{tabular}{c|c}
($\frac{1}{3}$, $\frac{2}{3}$, $\frac{3}{4}$) & ($\frac{2}{3}$, $\frac{1}{3}$,
$\frac{1}{4}$)\\[2pt]
\hline
($2u''$,$u''$,0) & ($2u''$,$u''$,0)\\
(0,-$u''$,$v''$) & (0,-$u''$,-$v''$)\\
(0,-$u''$,$v''$) & (0,$u''$,$v''$)\\
($2u''$,$u''$,0) & (-$2u''$,-$u''$,0)\\
\end{tabular}\\
\hline
\end{tabular}
\end{tabular}
\end{center}
\end{table*}

\section{Discussion and conclusions}
\label{sec_conclusions}
To summarize, we have reported high-resolution neutron powder
diffraction measurements in the orbitally-degenerate frustrated
triangular magnet 2H-AgNiO$_2$. We have observed a set of weak
structural reflections, undetected in previous x-ray measurements,
which indicate a small structural distortion with a tripling of
the unit cell in the hexagonal basal plane. We have proposed that
this could be explained by a periodic contraction and expansion of
NiO$_6$ octahedra in a three-sublattice structure as a consequence
of charge disproportionation on the Ni sites. We have also
observed that the triple cell peaks disappear at high temperatures
above $T_{S}=365$~K indicating a structural transition to the
ideal undistorted structure, where all Ni sites are identical
implying that charge is uniformly distributed on the Ni sites. The
low-temperature magnetic diffraction pattern is well explained by
a structure of ferromagnetic rows ordered antiferromagnetically,
but interestingly with ordered moments present only on the
electron-rich Ni sites, which as a consequence of charge
disproportionation would have the largest spin moment ($S=1$ if
Ni$^{2+}$). The observed magnetic structure on the ordered sites
has 2-fold symmetry compared to the 3-fold symmetry of the crystal
structure. We have proposed that the magnetic structure could be
explained starting from an easy-axis triangular lattice
antiferromagnet with additional weak in-plane next-nearest
neighbor antiferromagnetic couplings and/or weak ferromagnetic
interlayer interactions. Determination of the relative magnitude
of the exchanges and anisotropy terms requires measurements of the
spin gap and spin-wave dispersion band-width and such measurements
are in progress.\cite{wheeler}

Both the magnetic order and structural distortion observed here
are very different from the prevailing theoretical model for
Jahn-Teller active transition metal ions coupled in a triangular
lattice arrangement by near 90$^{\circ}$ metal-oxygen-metal bonds,
which predicts ferro-distortive orbital order and dominant
ferromagnetic in-plane interactions,\cite{khomskii} as indeed
observed experimentally in NaNiO$_2$ (Ref. \onlinecite{nanio2}).
2H-AgNiO$_2$ shows a different type of structural modulation which
suggests an alternative mechanism of lifting the large degeneracy
in the orbital sector by means of charge ordering, leading to
non-equivalent Ni sites, some electron rich and others electron
depleted. In the ideal, undistorted crystal structure each Ni site
has one electron in the two-fold degenerate $e_g$ orbital and at
the structural transition below $T_S=365$~K the lattice separates
into two sub-systems. From each hexagon of Ni ions one electron
jumps to the Ni site in the center to form an
orbitally-nondegenerate $e^2_g$ state, and this leaves a
surrounding honeycombe network of mainly electronically-inactive
Ni$^{4+}$ sites but with an extra electron for every two sites
(most likely itinerant and therefore distributed with equal
probability on every site, $1/8$ filled $e_g$ orbital on the
honeycombe network). Further experiments in particular NMR, X-ray
absorption (XAFS) or photoelectron spectroscopy (XPS) would be
needed to confirm such a scenario and determine quantitatively the
extent of the charge disproportionation.

Further studies are also needed to understand better the very
different magnetic behaviour of the expanded and contracted Ni
sites, attributed in the charge-order scenario to electron-rich
and -depleted sites, respectively. Band structure
calculations\cite{short_paper} suggest that as a result of charge
order the Ni ions in the center of expanded NiO$_6$ octahedra
become more localized and then the magnetic order at those sites
at low temperatures could be understood in terms of stacked
triangular lattices with a large spin moment (valence close to
Ni$^{2+}$ with $S=1$), whereas the remaining electron-depleted Ni2
and Ni3 sites located inside contracted NiO$_6$ octahedra maintain
a large itinerant character (due to shorter Ni-O distances and
thus stronger orbital overlap with the oxygens); in this scenario
the itinerant sites do not show a strong tendency to magnetic
order because of insufficiently large density of states at the
Fermi level. In fact, the band structure calculations suggested
that a small ordered moment of $\sim0.1~\mu_B$ may be induced on
the contracted Ni3 sites by the ordering of the large Ni1 moments.
We note that recent $\mu$SR measurements\cite{musr} have reported
an anomalous temperature-dependence of the local static magnetic
fields which may be due to small ordered moments on the honeycombe
Ni sites with a different temperature-dependence compared to the
large moments on the Ni1 sublattice.

Another interesting aspect worthwhile looking into is the role of
interlayer coupling in stabilizing the magnetic order and
structural distortion. It could be investigated by looking at a
system in which that coupling could be modified without modifying
the intralayer exchange integrals. AgNiO$_2$ is a rare example of
a delafossite having two polymorphs whose synthezis is feasible
\cite{polytype} and therefore the ideal candidate for further
studies seems to be the 3R-AgNiO$_2$ polytype. The only difference
between the latter and the 2H-AgNiO$_2$ is the way the NiO$_2$
layers are stacked on top of each other.
\section{Acknowledgements}
\label{sec_acknowledgements}
We would like to thank I.I. Mazin and M.D. Johannes for
collaboration on related work and and S.J. Blundell, T. Lancaster,
N. Shannon, and R. Moessner for useful discussions. We also thank
E. Suard and W. Kockelmann for technical assistance with the
experiments at the ILL and ISIS, D. Prabhakaran for help with the
SQUID and heat capacity measurements, and the initial stages of
sample preparation, and A. El-Turki for performing the SEM/EDX
analysis. The research was supported in part by EPSRC U.K. grants
EP/C51078X/2 (EW) and GR/R76714/02 (RC), a CASE award from the
EPSRC and ILL (EMW), the EU programme at the ILL. The D2B
measurements were supported by EPSRC U.K. grant GR/R88601/02.


\appendix
\section{Structure factors}
\label{app_A} Here we list the measured and fitted nuclear (Table\
\ref{tab_sf_nuclear}) and magnetic (Table\ \ref{tab_sf_mag})
structure factors.

\begin{table}[tbh]
\begin{center}
\caption{\label{tab_sf_nuclear} List of supercell structural peaks
associated with the tripling of the unit cell in the basal plane
with the observed and calculated unit cell structure factors for the
model of expanded and contracted Ni-O bonds shown in Fig.\
\ref{fig_structure}. The observed $|F|^2$ is the peak intensity
corrected for instrumental resolution effects, divided by the peak
multiplicity and normalized per unit cell of the P6$_3 2 2$ group.
For completeness a selection of the nominal peaks of the ideal
structure (P$6_3/mmc$ space group) is also given, and the peak
indices are given in both space groups.}
\begin{tabular}{c|c|c|c|c|c}
\hline
& $Q$         & $(h,k,l)$& $(h,k,l)$  & observed $|F|$ & calculated $|F|$\\
& (\AA$^{-1}$)& P$6_3 2 2$   & P$6_3/mmc$ &  (10$^{-14}$ m) & (10$^{-14}$ m) \\
\hline
\rotatebox[origin=lt]{90}{nominal peaks}
&
\begin{tabular}{c}
1.026\\
2.052\\
2.469\\
2.523\\
2.675\\
2.911\\
3.080\\
3.212\\
3.562\\
3.947\\
4.105\\
4.278\\
4.359\\
\end{tabular}
&
\begin{tabular}{c}
002\\
004\\
110\\
111\\
112\\
113\\
006\\
114\\
115\\
116\\
008\\
300\\
117\\
\end{tabular}
&
\begin{tabular}{c}
002\\
004\\
100\\
101\\
102\\
103\\
006\\
104\\
105\\
106\\
008\\
110\\
107\\
\end{tabular}
&
\begin{tabular}{c}
6.24\\
7.19\\
1.14\\
5.89\\
5.86\\
2.79\\
4.49\\
5.83\\
6.40\\
11.36\\
5.69\\
18.07\\
5.34\\
\end{tabular}
&
\begin{tabular}{c}
6.36\\
6.76\\
0.94\\
5.98\\
6.12\\
2.90\\
4.30\\
5.92\\
6.52\\
11.40\\
5.43\\
16.55\\
5.40\\
\end{tabular}\\
\hline
\rotatebox[origin=lt]{90}{supercell peaks}
&
\begin{tabular}{c}
1.425\\
1.515\\
1.756\\
2.097\\
2.851\\
2.898\\
2.936\\
3.031\\
3.393\\
3.514\\
3.807\\
3.836\\
3.865\\
3.910\\
4.074\\
4.105\\
4.295\\
4.562\\
4.586\\
4.998\\
5.166\\
5.209\\
5.242\\
5.366\\
5.535\\
5.575\\
5.703\\
5.726\\
5.745\\
5.794\\
5.821\\
5.870\\
\end{tabular}
&
\begin{tabular}{c}
010\\
011\\
012\\
013\\
020\\
021\\
015\\
022\\
016\\
024\\
211\\
025\\
017\\
212\\
213\\
026\\
214\\
215\\
027\\
028\\
311\\
217\\
312\\
313\\
314\\
218\\
040\\
041\\
315\\
042\\
0~1~11\\
0~2~10\\
\end{tabular}
&
\begin{tabular}{c}
2/3 -1/3 0\\
2/3 -1/3 1\\
2/3 -1/3 2\\
2/3 -1/3 3\\
4/3 -2/3 0\\
4/3 -2/3 1\\
2/3 -1/3 5\\
4/3 -2/3 2\\
2/3 -1/3 6\\
4/3 -2/3 4\\
4/3 1/3 1\\
4/3 -2/3 5\\
2/3 -1/3 7\\
4/3 1/3 2\\
4/3 1/3 3\\
4/3 -2/3 6\\
4/3 1/3 4\\
4/3 1/3 5\\
4/3 -2/3 7\\
4/3 -2/3 8\\
5/3 2/3 1\\
4/3 1/3 7\\
5/3 2/3 2\\
5/3 2/3 3\\
5/3 2/3 4\\
4/3 1/3 8\\
8/3 -4/3 0\\
8/3 -4/3 1\\
5/3 2/3 5\\
8/3 -4/3 2\\
2/3 -1/3 11\\
4/3 -2/3 10\\
\end{tabular}
&
\begin{tabular}{c}
0.15\\
0.24\\
0.05\\
0.02\\
0.31\\
0.57\\
0.23\\
0.14\\
0.00\\
0.00\\
0.35\\
0.53\\
0.46\\
0.60\\
0.37\\
0.39\\
0.73\\
0.30\\
0.49\\
0.35\\
0.67\\
0.37\\
0.71\\
0.56\\
0.77\\
0.63\\
0.54\\
0.89\\
0.56\\
0.41\\
0.22\\
0.17\\
\end{tabular}
&
\begin{tabular}{c}
0.16\\
0.24\\
0.09\\
0.02\\
0.35\\
0.53\\
0.23\\
0.19\\
0.16\\
0.15\\
0.22\\
0.50\\
0.26\\
0.62\\
0.42\\
0.35\\
0.66\\
0.26\\
0.56\\
0.22\\
0.67\\
0.19\\
0.74\\
0.48\\
0.77\\
0.58\\
0.60\\
0.90\\
0.65\\
0.32\\
0.21\\
0.12\\
\end{tabular}
\\
\hline
\end{tabular}
\end{center}
\end{table}

\begin{table}[htb]
\begin{center}
\caption{\label{tab_sf_mag} List of magnetic Bragg peaks with the
observed and calculated unit cell structure factors for the
magnetic structure in Fig.\ \ref{fig_mag_struct}. The observed
$|F|^2$ is the peak intensity corrected for instrumental
resolution effects, divided by the peak multiplicity and
normalized per unit cell of the P6$_3 2 2$ group. For completeness
the peak indices are given both in the distorted (P$6_3 2 2$) and
the ideal (P$6_3/mmc$) crystal structures.}
\begin{tabular}{c|c|c|c|c}
\hline
$Q$         & $(h,k,l)$& $(h,k,l)$  & observed $|F|$ & calculated $|F|$ \\
(\AA$^{-1}$)& P$6_3 2 2$   & P$6_3/mmc$ &  (10$^{-14}$ m) & (10$^{-14}$ m) \\
\hline
\begin{tabular}{c}
0.714\\
0.879\\
1.236\\
1.250\\
1.339\\
1.607\\
1.697\\
1.889\\
1.957\\
1.957\\
1.974\\
2.142\\
2.150\\
2.150\\
2.173\\
2.202\\
\end{tabular}
&
\begin{tabular}{c}
1/2 0 0 \\
1/2 0 1 \\
1/2 -1 0 \\
1/2 0 2 \\
1/2 -1 1 \\
1/2 -1 2 \\
1/2 0 3 \\
1/2 1 0 \\
1/2 1 1 \\
3/2 -1 1\\
1/2 -1 3\\
3/2 0 0 \\
1/2 1 2 \\
3/2 -1 2 \\
1/2 0 4 \\
3/2 0 1 \\
\end{tabular}
&
\begin{tabular}{c}
1/6 1/6 0\\
1/6 1/6 1\\
1/2 0 0 \\
1/6 1/6 2\\
1/2 0 1 \\
1/2 0 2\\
1/6 1/6 3\\
-1/6 5/6 0\\
1/6 5/6 1\\
1/6 5/6 1\\
1/2 0 3 \\
1/2 -1 0\\
1/6 5/6 2\\
1/6 5/6 2\\
1/6 1/6 4\\
1/2 -1 1\\
\end{tabular}
&
\begin{tabular}{c}
      0.61\\
      0.28\\
      0.00\\
      0.37\\
      0.74\\
      0.00\\
      0.16\\
      0.58\\
      0.34\\
      0.34\\
      0.48\\
      0.00\\
      0.50\\
      0.50\\
      0.15\\
      0.70\\
\end{tabular}
&
\begin{tabular}{c}
       0.71\\
       0.33\\
       0.00\\
       0.39\\
       0.72\\
       0.00\\
       0.16\\
       0.63\\
       0.35\\
       0.35\\
       0.45\\
       0.00\\
       0.53\\
       0.53\\
       0.20\\
       0.68\\
\end{tabular}\\
\hline
\end{tabular}
\end{center}
\end{table}

\section{Co-representation analysis of the triple-cell crystal
structure} \label{app_B}

Here we look in more detail at the symmetry properties of the
experimentally determined supercell modulation and show how the
the basis vectors for the charge and displacement order patterns
can be used to uniquely identify the distorted space group.
Specifically we will find all symmetry operations that leave the
order pattern invariant to determine the space group. We follow
closely ref. \onlinecite{Radaelli} where the basic concepts of
co-representation analysis as applied to the symmetry reduction
for $\bm{q}$-vector modulations at a generic point in the
Brillouin zone are explained. We first construct from the
irreducible representation (\emph{irrep}) modes a new set of
modes, known as corepresentation (\emph{corep}) modes, which are
invariant upon application of the anti-unitary operator $KI$,
where $I$ is the inversion operator ($h_{13}$ in Kovalev notation)
and $K$ is the complex conjugation. When combined with their
complex conjugate, these modes are centrosymmetric even if the
propagation vector is not equivalent to its inverse, as it is the
case here. In other words, the ``corep little group" contains all
24 symmetry operators in P$6_3/mmc$. We can then re-write both the
scalar and the polar vector modulation as a linear combinations of
corep modes, and directly assess their symmetry. It is important
to remember that the coefficients of the corep modes are
complex-conjugated upon application of an anti-unitary operator.

The irrep-corep matrix for space group 194 and propagation vector
${\bm q}_0=(1/3,1/3,0)$ is listed in Kovalev \cite{Kovalev}, and,
for irreducible representations $\tau_1$ and $\tau_2$ can be
summarized as follows:
\begin{description}
\item[-] Each of the two irreps generates exactly one corep,
which has the same matrices as the original irrep on the unitary operators.
For this reason, we will still use the notation $\tau_1$ and $\tau_2$ to indicate the
generated coreps.
\item[-]Operators $h_1$, $h_3$, $h_5$, $h_7$, $h_9$, $h_{11}$,
$Kh_{13}$, $Kh_{15}$, $Kh_{17}$, $Kh_{19}$, $Kh_{21}$, $Kh_{23}$ :
all coefficients are 1 for both $\tau_1$ and $\tau_2$
\item[-]Operators $h_{16}$, $h_{18}$, $h_{14}$, $h_{22}$,
$h_{24}$, $h_{20}$, $Kh_{4}$, $Kh_{6}$, $Kh_{2}$, $Kh_{10}$,
 $Kh_{12}$, $Kh_{8}$:  coefficients are 1 for $\tau_1$ and -1 for
 $\tau_2$
\end{description}

The corep modes $\delta$ can be obtained from the irrep modes
$\tau$ as
\begin{equation}
\label{corep_from_irrep} \delta=\frac{1}{2}(\tau+KI\tau)
\end{equation}

\subsection{Scalar Modes}

Since Ni is on the inversion center, and the mode is scalar and
real, we have $KI\tau=\tau$, and $\delta=\tau$. We can then
re-write the scalar modulation in Eq.\ (\ref{eq_zeta}) as
\begin{equation}
\zeta=e^{i\pi/3}\left[(\delta_1+\delta_2)+e^{i4\pi/3}(\delta_1-\delta_2)\right]
\end{equation}

\subsection{Vector Modes}

Here, site permutation, polar vector inversion and complex
conjugation all come into play, since
\begin{eqnarray}
\label{antiinversion}
KI\bm{\tau}(1)=-\overline{\bm{\tau}(2)}e^{i4\pi/3}\\
KI\bm{\tau}(3)=-\overline{\bm{\tau}(4)}e^{i4\pi/3}\nonumber
\end{eqnarray}
where $\bm{\tau}(1)$ is the mode on atom 1 etc. The irrep modes
are
\begin{eqnarray}
\label{irrep_modes}
\bm{\tau}(1)=\pm \bm{\tau}(3)=(e^{i\pi/6}, e^{i\pi/2},0)\\
\bm{\tau}(2)=\pm \bm{\tau}(4)=(e^{i\pi/2},e^{i\pi/6},0)\nonumber
\end{eqnarray}
where the $+$ and $-$ are for irrep 1 and 2, respectively (see
$\bm{\tau}_1$ and $\bm{\tau}_2$ in Table \ref{tab_po}). By
inserting eq.\ (\ref{antiinversion}) and (\ref{irrep_modes}) into
 eq.\ (\ref{corep_from_irrep}), and after some manipulation, one obtains
\begin{equation}
\bm{\delta}=\frac{\sqrt{3}}{2}e^{-i\pi/6}\bm{\tau}
\end{equation}
We can then re-write the vector modulation eq.\ (\ref{xi}) as
\begin{equation}
\bm{\xi}=\frac{\sqrt{3}}{3}e^{i4\pi/3}\left[(\bm{\delta}_1+\bm{\delta}_2)+e^{i4\pi/3}(\bm{\delta}_1-\bm{\delta}_2)\right]
\end{equation}

\subsection{Symmetry analysis}

It is now straightforward to perform the symmetry analysis on the
scalar and vector modulation. There are four cases, depending on
whether the operators are unitary or anti-unitary and whether
their matrices have the same or opposite signs for $\tau_1$ and
$\tau_2$ or $\bm{\tau}_1$ and $\bm{\tau}_2$.

\begin{description}

\item[\textbf{Case 1}]:  Unitary, same sign (e.g., $h_3$)

\begin{eqnarray}
h_3\zeta&=&e^{i\pi/3}\left[(\delta_1+\delta_2)+e^{i4\pi/3}(\delta_1-\delta_2)\right]=\zeta\\
h_3\bm{\xi}&=&\frac{\sqrt{3}}{3}e^{i4\pi/3}\left[(\bm{\delta}_1+\bm{\delta}_2)+e^{i4\pi/3}(\bm{\delta}_1-\bm{\delta}_2)\right]=\bm{\xi}\nonumber
\end{eqnarray}
so both modes are invariant.

\item[\textbf{Case 2}]:  Unitary, opposite sign (e.g., $h_{16}$)
\begin{eqnarray}
h_{16}\zeta&=&e^{i\pi/3}\left[(\delta_1-\delta_2)+e^{i4\pi/3}(\delta_1+\delta_2)\right]\ne \zeta\\
h_{16}\bm{\xi}&=&\frac{\sqrt{3}}{3}e^{i4\pi/3}\left[(\bm{\delta}_1-\bm{\delta}_2)+e^{i4\pi/3}(\bm{\delta}_1+\bm{\delta}_2)\right]
\ne \bm{\xi}\nonumber
\end{eqnarray}
so neither mode is invariant.

\item[\textbf{Case 3}]:  Antiunitary, same sign (e.g., $Kh_{13}$)
\begin{eqnarray}
Kh_{13}\zeta&=&e^{-i\pi/3}\left[(\delta_1+\delta_2)+e^{-i4\pi/3}(\delta_1-\delta_2)\right]\ne \zeta\\
Kh_{13}\bm{\xi}&=&\frac{\sqrt{3}}{3}e^{-i4\pi/3}\left[(\bm{\delta}_1+\bm{\delta}_2)+e^{-i4\pi/3}(\bm{\delta}_1-\bm{\delta}_2)\right]
\ne \bm{\xi}\nonumber
\end{eqnarray}
so neither mode is invariant.

\item[\textbf{Case 4}]:  Antiunitary, opposite sign (e.g.,
$Kh_{6}$)
\begin{eqnarray}
Kh_{6}\zeta&=&e^{-i\pi/3}\left[(\delta_1-\delta_2)+e^{-i4\pi/3}(\delta_1+\delta_2)\right]\\
&=&e^{-i5\pi/3}(\delta_1+\delta_2)+e^{-i\pi/3}(\delta_1-\delta_2) \nonumber \\
&=&e^{i\pi/3}(\delta_1+\delta_2)+e^{i5\pi/3}(\delta_1-\delta_2)=\zeta \nonumber \\
Kh_{6}\bm{\xi}&=&\frac{\sqrt{3}}{3}e^{-i4\pi/3}\left[(\bm{\delta}_1-\bm{\delta}_2)+e^{-i4\pi/3}(\bm{\delta}_1+\bm{\delta}_2)\right] \nonumber\\
&=&\frac{\sqrt{3}}{3}\left[e^{-i2\pi/3}(\bm{\delta}_1+\bm{\delta}_2)+e^{-i4\pi/3}(\bm{\delta}_1-\bm{\delta}_2)\right]\nonumber \\
&=&\frac{\sqrt{3}}{3}\left[e^{i4\pi/3}(\bm{\delta}_1+\bm{\delta}_2)+e^{i2\pi/3}(\bm{\delta}_1-\bm{\delta}_2)\right]=\bm{\xi}\nonumber
\end{eqnarray}
so both modes are invariant.

\end{description}

In summary, both modes are invariant upon application of the
operators $h_1$, $h_3$, $h_5$, $h_7$, $h_9$, $h_{11}$, $Kh_{4}$,
$Kh_{6}$, $Kh_{2}$, $Kh_{10}$, $Kh_{12}$, $Kh_{8}$ and the
resulting modulated structure is invariant by the same operators
\emph{without} the complex conjugation.  These operators represent
all the \emph{proper} rotations of space group $P6_3/mmc$ - in
other words, the six-fold screw axis and the associated orthogonal
2-fold axes.  Taking into account the loss of translational
symmetry due to the propagation vector, the resulting space group
is $P6_322$ with the unit cell of dimensions
$\sqrt{3}a_0\times\sqrt{3}a_0\times c$.



\end{document}